\documentclass[aps,prl,twocolumn,groupedaddress,dvips]{revtex4}
\usepackage[dvips]{graphicx}
\begin{document}
\preprint{}
\title{Saturation of Spin-Polarized Current in Nanometer Scale Aluminum Grains}

\author{Y. G. Wei, C. E. Malec, D. Davidovi\'c }
\affiliation{School of Physics, Georgia Institute of Technology,
Atlanta, GA 30332}
\date{\today}
\begin{abstract}
We describe measurements of spin-polarized tunnelling via discrete
energy levels of single Aluminum grains. In high resistance
samples ($\sim G\Omega$), the spin-polarized tunnelling current
rapidly saturates as a function of the bias voltage. This
indicates that spin-polarized current is carried only via the
ground state and the few lowest in energy excited states of the
grain. At the saturation voltage, the spin-relaxation rate
$T_1^{-1}$ of the highest excited states is comparable to the
electron tunnelling rate: $T_1^{-1}\approx 1.5\cdot 10^6 s^{-1}$
and $10^7s^{-1}$ in two samples. The ratio of $T_1^{-1}$ to the
electron-phonon relaxation rate is in agreement with the
Elliot-Yafet scaling, an evidence that spin-relaxation in Al
grains is governed by the spin-orbit interaction.

\end{abstract}

\maketitle

\section{introduction}

Electron tunnelling through single nanometer scale metallic grains
at low temperatures can display a discrete energy level
spectrum.~\cite{ralph} Tunnelling spectroscopy of the energy
spectra have led to numerous discoveries, including Fermi-Liquid
coupling constants between quasiparticles,~\cite{agam} spin-orbit
interactions,~\cite{davidovic,petta} and superconducting
correlations in zero-dimensional systems.~\cite{black} Some
information regarding the spin of an electron occupying a discrete
level can be obtained using spin-unpolarized tunnelling, such as
spin-multiplicity and electron g-factors.~\cite{ralph}

In this letter we report on spin-polarized tunnelling via discrete
energy levels of single aluminum grains. Spin-polarized electron
transport permits studies of spin relaxation and spin
dephasing.~\cite{johnson,jedema1} By comparison, spin-unpolarized
spectroscopy is suitable for the studies of energy relaxation in
the grains.~\cite{agam,ralph} Since spin-relaxation times are
generally many orders of magnitude longer than energy relaxation
times, spin-unpolarized spectroscopy is not an easy tool to study
spin-relaxation in the grains and spin-polarized tunnelling is
needed. We find that some electron spin-relaxation times in Al
grains are exceptionally long compared to bulk Al with similar
disorder, on the order of $\mu s$.

Spin-polarized transport via metallic grains has recently
generated a lot of theoretical
interest.~\cite{braun,weymann,vandermolen,wetzels,cottet1} In
addition, there is a major effort to study nano-spintronics using
carbon-nanotubes; see Ref.~\cite{cottet} and references therein.
Spin-coherent electron tunnelling via nanometer scale normal
metallic grains has been confirmed in arrays~\cite{zhang1,ernult}
and in single grains.~\cite{bernard} However, the electron
spin-relaxation time $T_1$ in a metallic grain has not been
reported yet.

\section{Sample fabrication}

\begin{figure}
\includegraphics[width=0.45\textwidth]{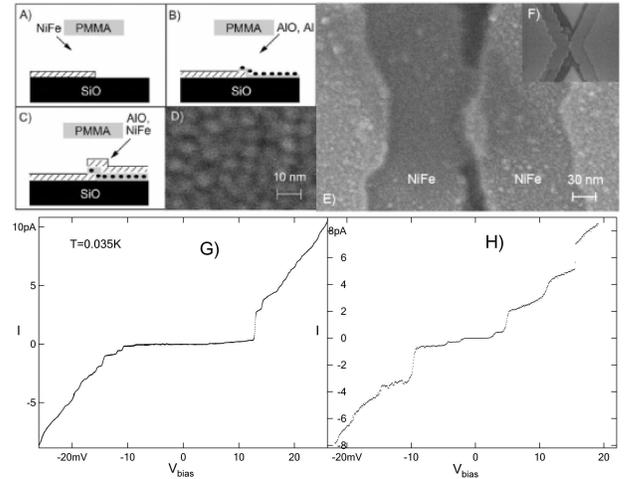}
\caption{A, B, C: Sample fabrication steps. D. Image of Al grains.
E, F: Image of a typical sample. G, H: I-V curves at the base
temperature.~\label{fig1}}
\end{figure}

Our samples are prepared by electron beam lithography and shadow
evaporation, similar to the technique described
previously.~\cite{davidovic} First we define a resist bridge
placed 250 nm above the Si wafer; this bridge acts as a mask. Next
(Fig.~\ref{fig1}-A), we deposit 11 nm permalloy (Py =
Ni$_{0.8}$Fe$_{0.2}$) onto oxidized silicon substrate at $4\cdot
10^{-7}$ Torr base pressure, measured near the gate valve, along
the direction indicated by the arrow. Then we rotate the sample by
36 degrees without breaking the vacuum and deposit 1.2 nm of
Al$_2$O$_3$ by reactive evaporation of Al,~\cite{davidovic} at a
rate of 0.35 nm/s, at an oxygen pressure of $2.5\cdot 10^{-5}$
Torr. Now, oxygen flow is shut down. When pressure decreases to
the $10^{-7}$ Torr range, we deposit a 0.6 nm thick film of Al, as
shown in Fig.~\ref{fig1}-B. Al forms isolated grains with a
typical diameter of 5 nm. The grains are displayed by the scanning
electron microscope (SEM) image in Fig.~\ref{fig1}-D. Finally we
deposit another 1.2 nm layer of Al$_2$O$_3$ by the reactive
evaporation and top it of by an 11 nm thick film of Py
(Fig.~\ref{fig1}-C).
We make many samples on the same silicon wafer, and vary the
overlap from 0 to 50 nm and select the devices with the highest
resistance, as they have the smallest overlap. Figs.~\ref{fig1}-E
and F show SEM images of a typical device.

\section{Discrete energy levels}

Transport properties of the samples at low temperatures were
measured using an Ithaco current amplifier. The samples were
cooled down to $\approx 0.035K$ base temperature. The sample leads
were cryogenically filtered to reduce the electron temperature
down to $\approx 0.1K$.

The majority of samples ($>$80\%) exhibit Coulomb Blockade at low
temperature. About 150 samples were measured at 4.2K and 16
samples at $0.035 K$. In this paper we describe two samples. The
I-V curve of two samples are shown in Fig.~\ref{fig1}-G and H. The
tunnelling current increases in discrete steps as a function of
bias voltage, corresponding to discrete electron-in-a-box energy
levels of the grain.

In sample 1, the average electron-in-a-box level spacing caused by
electron geometric confinement is $\delta\approx 0.8meV$, which
corresponds to diameter of $D\approx 6nm$ assuming a spherical Al
grain. The average current step is ${\overline I}\approx 0.47pA$.
We make a connection with the tunnelling rates from the leads to
the grain and the measured current response.  The tunnel junctions
are highly asymmetric, and therefore one of the tunnelling rates
is much smaller than the other, and thus rate limiting. Throughout
this paper, we choose the rate limiting step to be across the left
junction, corresponding to the tunnelling rate $\Gamma_L$.
Therefore, our measured current corresponds to the average
tunnelling-in rate of $\overline{\Gamma}_L={\overline
I}/2|e|\approx 1.5\cdot 10^6 s^{-1}$. Similarly, in sample 2,
$\delta\approx 2.7 meV$, $D\approx 4nm$, and
${\overline\Gamma_L}\approx 9.6\cdot 10^6 s^{-1}$.

 The
spin-conserving energy-relaxation in Al grains takes place by
phonon emission with the relaxation rate~\cite{agam}
\begin{equation}
{\tau_{e-ph}^{-1}(\omega)}=\left(\frac{2}{3}E_F\right)^2\frac{\omega^3\tau_e\delta}{2\rho\hbar^5
v_S^5}, \label{eph}
\end{equation}
where $E_F=11.7 eV$ is the Fermi energy, $\omega$ is the energy
difference between the initial and the final state, $\rho=2.7
g/cm^3$ is the ion-mass density, and $v_s=6420 m/s$ is the sound
velocity. We obtain $\tau_{e-ph}^{-1}(\delta)\approx 1.6\cdot 10^9
s^{-1}$ and $4.1\cdot 10^{10} s^{-1}$ in samples 1 and 2,
respectively. Sample 2 has significantly larger relaxation rate
because of the larger level spacing. Since the tunnelling rates in
our samples are $\sim 10^6s^{-1}$, if the grain is excited by
electron tunnelling in and out, it will instantly relax to the
lowest energy state accessible by spin-conserving transitions.

\begin{figure}
\includegraphics[width=0.45\textwidth]{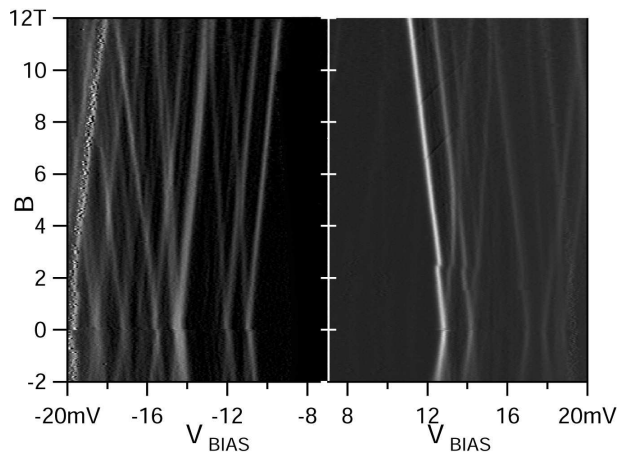} \caption{A, B: Differential conductrance (gray) versus bias
voltage and the applied magnetic field in sample 1 at the base
temperature.~\label{fig2}}
\end{figure}

As shown by Fig.~\ref{fig2}, the energy levels exhibit Zeeman
splitting as a function of an applied magnetic field. In sample 1,
the I-V curve probes the same energy spectrum at  negative and
positive bias voltage. This is evident from the equivalence of the
magnetic field dependencies at negative and positive bias. The
lowest tunnelling threshold is two fold degenerate at zero
magnetic field, showing that $N_0$, the number of electrons on the
grain before tunnelling in, is even. The conductance peaks are
similar in magnitude at negative bias, because the first
tunnelling step, in which an electron tunnels in to the grain
through the higher resistance junction, is rate limiting. At
positive bias, the first conductance peak is much larger than the
subsequent conductance peaks, because the first tunnelling step
takes place via the lower resistance junction, and the rates are
limited by the electron discharge process across the high
resistance junction.

In sample 1, the first two peaks split corresponding to g-factors:
$g=1.83\pm 0.05$ and $1.95\pm 0.05$. Slight reduction of the
g-factors from $2$ indicates spin-orbit interaction in
Al.~\cite{ralph} The avoided level crossings  clearly are resolved
in Fig.~\ref{fig2}, near points $(-11.5mV,5T)$ and
$(-13mV,11.5T)$. The corresponding avoided crossings at positive
bias are located near $(13.5mV,5T)$ and $(15.5mV,11.5T)$,
respectively. In the regime where g factors are slightly reduced,
the spin-orbit scattering rate ($\tau_{SO}^{-1}$) can be obtained
from the avoided crossing energies $\Delta_{SO}\approx
0.1meV$.~\cite{adam} Theory predicts that
$\tau_{SO}\approx\hbar\delta/\pi\Delta_{SO}^2$,~\cite{adam} within
a factor of two. Thus, we obtain $\tau_{SO}^{-1}\approx 5.5\cdot
10^{10} s^{-1}$. By the Elliot-Yafet relation,~\cite{yafet}
$\tau_{SO}^{-1}$ is related to the elastic scattering rate
$\tau_e^{-1}$: $\tau_{SO}^{-1}=\alpha \tau_e^{-1}$. Assuming
ballistic grain, $\tau_{e}^{-1}\approx v_F/D=3.4\cdot 10^{14}
sec^{-1}$. We obtain $\alpha\approx 1.6\cdot 10^{-4}$, in
excellent agreement with $\alpha\approx 10^{-4}$ in Al thin
films.~\cite{jedema3}

\section{Spin-polarized tunnelling}

Now we discuss magnetoresistance from the spin-polarized
tunnelling. In the magnetic field range of $\pm 50 mT$,
approximately 90\% of the samples do not display any of the
tunnelling magnetoresistance effect (TMR) . By contrast, we tested
about 10 tunnelling junctions without the embedded grains and with
similar resistance (empty junctions) at 4.2 K. All of the empty
junctions exhibit a significant TMR in this field range,
comparable to $10\%$. Approximately one half of the empty
junctions display a simple spin-valve effect. So, the absence of
TMR for electron tunnelling via grains shows that the
spin-dephasing rate $T_2^{-1}$ in 90\% of the samples must be much
larger than the tunnelling rate.

Nevertheless, approximately 10\% of the samples with embedded
grains display significant TMR, so the dephasing must be weak,
e.g. $T_2^{-1}$ must be smaller than or comparable to the
tunnelling rate in these samples. $T_2$ variation among different
samples could be explained by magnetic defects, such as
paramagnetic impurities from the Py layer. Paramagnetic impurities
are common sources of dephasing.~\cite{bergman} The defects would
be located on the grain surface, since bulk Al does not support
paramagnetism. Since the number of atoms on the surface is
relatively small ($\sim$1000), we could occasionally obtain a
sample free of impurities.  More insight into the nature of $T_2$
in this device will require a more in depth theoretical study.

A majority of the samples with nonzero TMR show positive TMR near
the Coulomb-Blockade conduction threshold; only about 30\% of the
samples show negative TMR. The sign of TMR in quantum dots is
determined by the interplay between charging effects and
spin-accumulation.~\cite{barnas,ernult} For any given sample, the
data in this paper correspond to the voltage range within the
first step of the Coulomb staircase. In this range the sign of TMR
is found to be constant as expected.

TMR in our devices usually does not display a simple spin-valve
effect. We believe this is because there are spin-dependent
interactions inside the grain that induce a complicated TMR even
when the magnetic transitions in the drain and source leads are
sharp and as expected. For example, a rotation of stray magnetic
field acting on the grain will alter the direction of the
spin-quantization axis in the grain, thereby changing the
conductance~\cite{braun}. A rotation or a switch of a remote
domain can change the tunnelling current through the grain via the
magnetic field generated by the domain. Similarly, the orientation
of the nuclear spin in the grain can change the quantization axes
via the hyperfine interaction.
\begin{figure}
\includegraphics[width=0.45\textwidth]{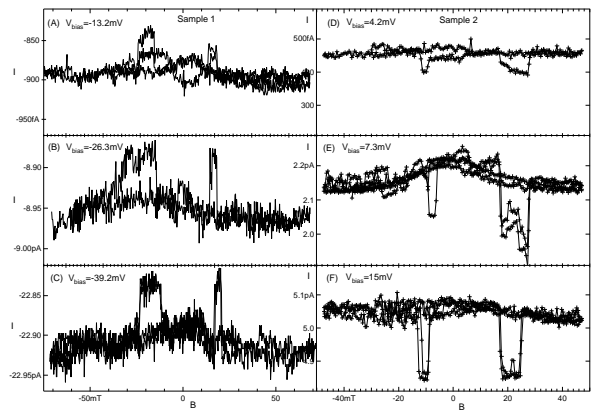} \caption{A-F: Spin-valve effect in current versus
applied magnetic field in two samples at the base temperature. The
current magnitude is reduced in the antiparallel
state.~\label{fig3}}
\end{figure}

We select only those samples that display a simple spin-valve TMR
effect, which is shown in Figs.~\ref{fig3}. Fig.~\ref{fig3}-A is
the TMR of sample 1 at a bias voltage corresponding to the second
current plateau. TMR is barely resolved in this case, since the
current changes by only about 40fA. We do not have good data to
display TMR at the first current plateau. By comparison,
Figs.~\ref{fig3}-B and C display TMR at bias voltage where the
number of electron-in-a-box levels energetically available for
tunnelling-in are approximately 19 and 48, respectively. To
facilitate comparisons, the current intervals on the vertical axes
in Figs.~\ref{fig3} A-C and D-F have equal lengths.

The main observation in this letter is that $\Delta
I=I_{\uparrow\uparrow}-I_{\uparrow\downarrow}$ is nearly constant
with current above a certain current. There is hardly any increase
in $\Delta I$ between Figs.~\ref{fig3} B and C and between
Figs.~\ref{fig3} E and F. This behavior is shown in more detail
Fig.~\ref{fig4}-A and B, which displays $\Delta I$ versus bias
voltage. $\Delta I$ versus negative bias voltage in sample 1 is
fully saturated at the third current plateau; at the second
current plateau, $\Delta I$ is already at one half of the
saturation value. Similarly, in sample 2 $\Delta I$ reaches
saturation at the second current plateau. Our samples should be
contrasted with ordinary ferromagnetic tunnelling junctions, where
$\Delta I$ is  proportional to the current over a significantly
wider range of bias voltage~\cite{moodera1,zhang2}.

\section{Interpretation of the results}

In coulomb-blockade samples containing magnetic leads, the
electrochemical potential difference between the island and leads
can jump when the magnetization in one of the leads changes
direction~\cite{vandermolen}. This can lead to a sudden shift in
energy levels, producing a jump in current that is constant as a
function of bias voltage. The shift in energy levels is seen as a
discontinuity near zero magnetic field in Fig. 2, and is $\sim 0.1
mV$.

To show that the electrocemical shift is not responsible for the
saturation of the spin-polarized current with voltage in our
sample, we performed other measurements by sweeping the magnetic
field both on and between the current plateaus, coming up with
similar values for the electrochemical shift.  The shift is lower
than the average level spacing of 0.8 meV and 2.7 meV for sample 1
and sample 2 respectively.  Therefore, since we measured
magnetoresistance in the middle of the current plateau, the
threshold voltage shift should not effect our measurements of the
saturation in $\Delta I$.

To explain $I_{\uparrow\uparrow}-I_{\uparrow\downarrow}=const$, we
must discuss the relative magnitudes of three rates:
$\tau_{e-ph}^{-1}$, the rate of energy relaxation from excited to
lower energy states by spin-conserving phonon emission;
$\Gamma_L$, the rate electrons tunnel into the grain; and
$T_1^{-1}$, the rate of transitions between levels that result in
an electron flipping its spin orientation. $\tau_{e-ph}^{-1}$ is
obtained theoretically, the measured I-V spectrum fixes the
tunnelling rate, and $T_1^{-1}$ is obtained from the saturation in
$I_{\uparrow\uparrow}-I_{\uparrow\downarrow}$ with bias voltage.

\begin{figure}
\includegraphics[width=0.45\textwidth]{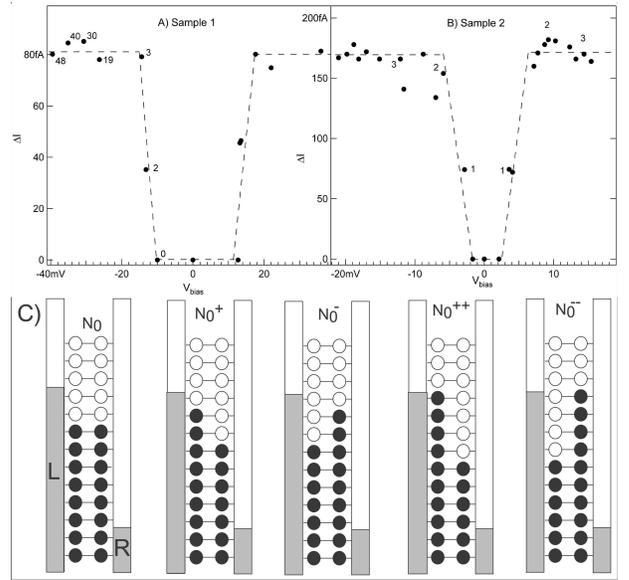}
\caption{A and B: $\Delta
I=|I_{\uparrow\uparrow}-I_{\uparrow\downarrow}|$ versus bias
voltage in samples 1 and 2, respectively, at the base temperature.
The numbers near the circles indicate how many doubly degenerate
electron-in-a-box levels are available for tunnelling in. C:
Possible spin-polarized electron configurations caused by electron
tunnelling in and out, before an electron tunnels in, at the
second current plateau, for $N_0$ even.~\label{fig4}}
\end{figure}

Finally we must deduce the relative magnitude of $T_1^{-1}$. The
rate of spin-flip transitions is expected to be significantly
smaller than $\tau_{e-ph}^{-1}$~\cite{yafet}. In this case the
ground state would not necessarily be accessible by energy
relaxation. The grain could remain in an excited, spin-polarized
state, as sketched in Fig.~\ref{fig4}-C. These spin-polarized
excited states are responsible for spin accumulation in the
antiparallel magnetic configuration of the leads. If the
relaxation rates for the spin-flip transitions are much smaller
than the tunnelling rate, then various spin-polarized states would
have similar probabilities, which are determined by the tunnelling
rates. In the antiparallel configuration of the leads, the
probabilities of the excitations with spin up would be enhanced by
$1+ P$ and probabilities of the excitations with spin down would
be suppressed by $1-P$, where $P$ is the spin-polarization in the
leads. In the parallel configurations, the probabilities of the
excitations with spin up and spin down are the same. In this
regime, $I_{\uparrow\uparrow}-I_{\uparrow\downarrow}$ is
proportional to the current, similar to the usual ferromagnetic
tunnelling junctions.

It is reasonable to expect that the spin-flip rate
$T_1^{-1}(\omega)$ increases rapidly with energy difference
$\omega$ between the initial and the final state~\footnote{In bulk
metals the spin-orbit scattering rate increases rapidly with
electron excitation energy}. If $T_1^{-1}(\omega)$ exceeds the
tunnelling rate above some $\omega$, then the excitations with
energy $>\omega$ will occur with a reduced probability in the
ensemble of states generated by tunnelling in and out. Thus
$\Delta I$ is limited by tunnelling via the ground state and those
low lying spin-polarized states where $T_1^{-1}(\omega)<\Gamma_L$.
$\Delta I$ versus bias voltage approaches saturation approximately
when $T_1^{-1}(\omega)=\Gamma_L$, where $\omega$ is the highest
excitation energy in the ensemble of spin-polarized states
generated by tunnelling in and out: $\omega\approx\delta
\frac{I}{|e|\Gamma_L}$. This is how we determine the
spin-relaxation time $T_1(\omega)$ at an energy $\omega$ in a
given sample.

In sample 1, $\Delta I$ is at 50\% of the saturation value at the
second current plateau, and $\Delta I$ is saturated at the third
current plateau. At the second current plateau, the
spin-relaxation rate of the highest energy excited state generated
by tunnelling must be close to the tunnelling rate.  Since the
spin relaxation is very rapid in configurations more than
$3\delta$ above the ground state, and $N_0$ is even as noted
above, the grain spends most of the time among the five
configurations shown in Fig.~\ref{fig4}-C: $N_0$, $N_0^+$,
$N_0^-$, $N_0^{++}$, and $N_0^{--}$. The highest energy
spin-polarized states are $N_0^{++}$ and $N_0^{--}$. Thus,
$T_1^{-1}(3\delta)\approx \Gamma_L=1.5\cdot 10^6 s^{-1}$. In
sample 2, this analysis leads to $T_1^{-1}(2\delta)\approx 10^7
s^{-1}$.

Now we discuss the origin of spin relaxation and its rapid
enhancement with the energy difference. Note that the rate of
spin-conserving transitions in Eq.~\ref{eph} increases as
$\omega^3$. We suggest that the electron-phonon transition rates
without and with spin flip scale by the Elliot-Yafet relation:
$T_1^{-1}(\omega)=\alpha'\tau_{e-ph}^{-1}(\omega)$. This scaling
would certainly explain the rapid increase in spin-relaxation rate
with excitation energy. In metallic films, it is well established
that the Elliot-Yafet scaling applies for both elastic and
inelastic scattering processes, with $\alpha\approx
\alpha'$.~\cite{jedema3}

In sample 1, Eq.~\ref{eph} leads to
$\tau_{e-ph}^{-1}(3\delta)\approx 4\cdot 10^{10} s^{-1}$. Since
$T_1^{-1}(3\delta)\approx 1.5\cdot 10^6 s^{-1}$, we obtain
$\alpha'\approx 0.4\cdot 10^{-4}$. Similarly, in sample 2,
$\tau_{e-ph}^{-1}(2\delta)\approx 3.3\cdot 10^{11} s^{-1}$ and we
obtain $\alpha'\approx 0.3\cdot 10^{-4}$. $\alpha'$ agrees with
$\alpha\approx 1.5\cdot 10^{-4}$ obtained earlier, within an order
of magnitude. So the ratio of $\tau_{e-ph}$ and $T_1$ is in
agreement with the Elliot-Yafet scaling. This is an evidence that
the spin-flip transitions in Al grains are driven by the
spin-orbit interaction. By this relaxation mechanism, the spin of
an electron on the grain is coupled to the phonon continuum via
the spin-orbit interaction. An electron in an excited
spin-polarized state relaxes by an emission of a phonon, which has
an angular momentum equal to the difference between the initial
and final electron spin.

\section{Conclusion}

In summary, we have observed spin-coherent electron transport via
discrete energy levels of single Al grains. Spin polarized current
saturates quickly as a function of bias voltage, which
demonstrates that the ground state and the lowest excited states
carry spin polarized current. Higher excited states have a
relaxation time shorter than the tunnelling time and they do not
carry spin-polarized current. The spin-relaxation time of the
low-lying excited states is $T_1\approx 0.7\mu s$ and $0.1\mu s $
in two samples. Finally, the ratio of the spin-flip transition
rate and the electron-phonon relaxation rate is in quantitative
agreement with the Elliot-Yafet scaling ratio, an evidence that
the spin-relaxation transitions are driven by the spin-orbit
interaction.

This work was performed in part at the Georgia-Tech electron
microscopy facility. We thank Matthias Braun and Markus Kindermann
for consultation. This research is supported by the DOE grant
DE-FG02-06ER46281 and David and Lucile Packard Foundation grant
2000-13874.

\bibliography{career1}

\end{document}